\documentclass[aps,prl,reprint,superscriptaddress,nobalancelastpage]{revtex4-2}
\usepackage{graphicx}
\usepackage{xcolor}
\usepackage{amsmath}
\usepackage{ulem}
\usepackage{bm}
\DeclareMathOperator{\sech}{sech}
\usepackage{balance}
\begin{document}


\title{{Anharmonic phonon coupling enabled by local inversion symmetry breaking at domain walls in ferroelastics}}



\author{Seyyed Jabbar Mousavi}
\thanks{These authors contributed equally to this work.}
\email[]{Contact author: jabbar.mousavi@unibe.ch}
\affiliation{\mbox{Institute of Applied Physics, University of Bern, Sidlerstrasse 5, 3012 Bern, Switzerland}}
\author{Vivek Unikandanunni}
\thanks{These authors contributed equally to this work.}
\affiliation{\mbox{Institute of Applied Physics, University of Bern, Sidlerstrasse 5, 3012 Bern, Switzerland}}
\author{Niccolò Sellati}
\thanks{These authors contributed equally to this work.}
\affiliation{\mbox{Department of Physics, Sapienza University of Rome, P.le A. Moro 5, 00185 Rome, Italy}}
\author{Paolo Barone}
\affiliation{\mbox{Consiglio Nazionale delle Ricerche CNR-SPIN, Area della Ricerca di Tor Vergata, 00133, Rome, Italy}}
\author{Martina Basini}
\affiliation{\mbox{Physics Department, ETH Zurich, Zurich, Switzerland}}
\author{Steven L. Johnson}
\affiliation{\mbox{Physics Department, ETH Zurich, Zurich, Switzerland}}
\author{Andrey Shalit}
\affiliation{\mbox{Department of Chemistry, University of Zurich, Winterthurerstrasse 190, CH-8057 Zurich, Switzerland}}
\author{Peter Hamm}
\affiliation{\mbox{Department of Chemistry, University of Zurich, Winterthurerstrasse 190, CH-8057 Zurich, Switzerland}}
\author{Mattia Udina}
\affiliation{\mbox{Laboratoire Matériaux et Phénomènes Quantiques, Université Paris Cité, CNRS, 75205 Paris, France}}
\affiliation{Institut de Physique et Chimie des Matériaux de Strasbourg (UMR 7504), Université de Strasbourg and CNRS, Strasbourg, 67200, France}
\author{Thomas Feurer}
\affiliation{\mbox{Institute of Applied Physics, University of Bern, Sidlerstrasse 5, 3012 Bern, Switzerland}}


\date{\today}

\begin{abstract}
In ferroelastic materials, spontaneous symmetry breaking leads to the formation of twin domains. Although the bulk crystal typically remains centrosymmetric, inversion symmetry can be locally broken at the domain walls, potentially changing phonon selection rules and enabling local anharmonic phonon coupling. Here we report direct evidence of such anharmonic coupling in ferroelastic LaAlO$_3$ using two-dimensional Raman-terahertz spectroscopy. We attribute the cross-peaks observed in the two-dimensional spectra to both mechanical and electrical anharmonicity between the $A_{1g}$ Raman-active phonon and the $E_g$ phonon, which acquires finite infrared activity through local inversion symmetry breaking at ferroelastic domain walls. These findings provide new insight into the complex lattice dynamics of ferroelastic materials and highlight the potential of two-dimensional Raman-terahertz spectroscopy to uncover subtle symmetry breaking through the detection of intrinsically weak anharmonic signals.

\end{abstract}


\maketitle



\textit{Introduction}— 
Ferroelastic materials are characterized by a spontaneous strain that emerges at a structural phase transition and acts as an order parameter \cite{salje2012ferroelastic,nataf2020domain}. In contrast to ferromagnets and ferroelectrics, centrosymmetric ferroelastics preserve both time-reversal and inversion symmetry. This transition involves a symmetry-lowering lattice distortion without generating macroscopic polarization or magnetization. The distorted phase supports symmetry-equivalent orientation variants of the strain state that organize into ferroelastic twin domains \cite{arlt1990twinning,bueble1998influence,guo2019nanoscale,roeper2026uniaxial}.

A prototypical ferroelastic material is lanthanum aluminate LaAlO$_3$ (LAO) \cite{kustov2018laalo3}, which undergoes a cubic-to-rhombohedral transition at 813 K driven by an antiferrodistortive rotation of the oxygen octahedra. The resulting ferroelastic phase hosts soft phonon branches, in particular the lowest-lying $E_g$ and $A_{1g}$ modes associated with the octahedral tilt pattern \cite{hayward2002order,Hayward_2005,Abrashev_1999,basini2024terahertz,liu1995impulsive}. According to bulk crystal symmetry, these modes are Raman active but infrared inactive. However, the ferroelastic domain structure introduces additional symmetry considerations. While the bulk crystal remains centrosymmetric, ferroelastic domain walls (DWs) break translational symmetry and locally relax inversion symmetry at the interface \cite{yokota2018symmetry,yokota2021optical,yokota2020domain,salje2016direct}. Consequently, phonon selection rules can be locally modified, which gives rise in principle to weak but finite infrared activity of otherwise Raman-only modes and may enable subtle anharmonic phonon-phonon interactions localized at the DWs. Exploring these additional anharmonic pathways is crucial for understanding the complex lattice dynamics of LAO and other ferroelastic compounds \cite{suda2009anharmonicity,hortensius2020ultrafast,Neugebauer_2021}.

In this work we investigate anharmonic phonon-phonon interactions by impulsively exciting, with femtosecond optical pulses, the soft Raman-active phonons associated with the structural instability in LAO, and probing the resulting lattice dynamics with broadband terahertz radiation in a hybrid two-dimensional Raman-THz-THz (2D-RTT) \cite{Ciardi_2019, Shalit_2021, Mousavi_2022} spectroscopy scheme. This approach enables the direct detection of anharmonic excitation of infrared-active phonons driven by coherent Raman modes, thereby revealing the previously overlooked reciprocal counterpart of ionic Raman scattering (IRS) \cite{Maradudin_1970, Wallis_1971, Humphreys_1972, forst_2011, Subedi_2014, von_2018} and infrared-resonant Raman scattering (IRRS) \cite{khalsa2021,blank2023two}. 
Our measurements, supported by numerical simulations, demonstrate anharmonic couplings between the soft $A_{1g}$ and $E_g$ modes. Density functional theory (DFT) calculations reveal weak infrared activity of the $E_g$ phonon at ferroelastic domain walls, where local inversion symmetry is broken. This symmetry reduction enables anharmonic couplings that are forbidden in a translationally invariant bulk crystal.
\begin{figure}[t]
		\includegraphics[width=0.48\textwidth]{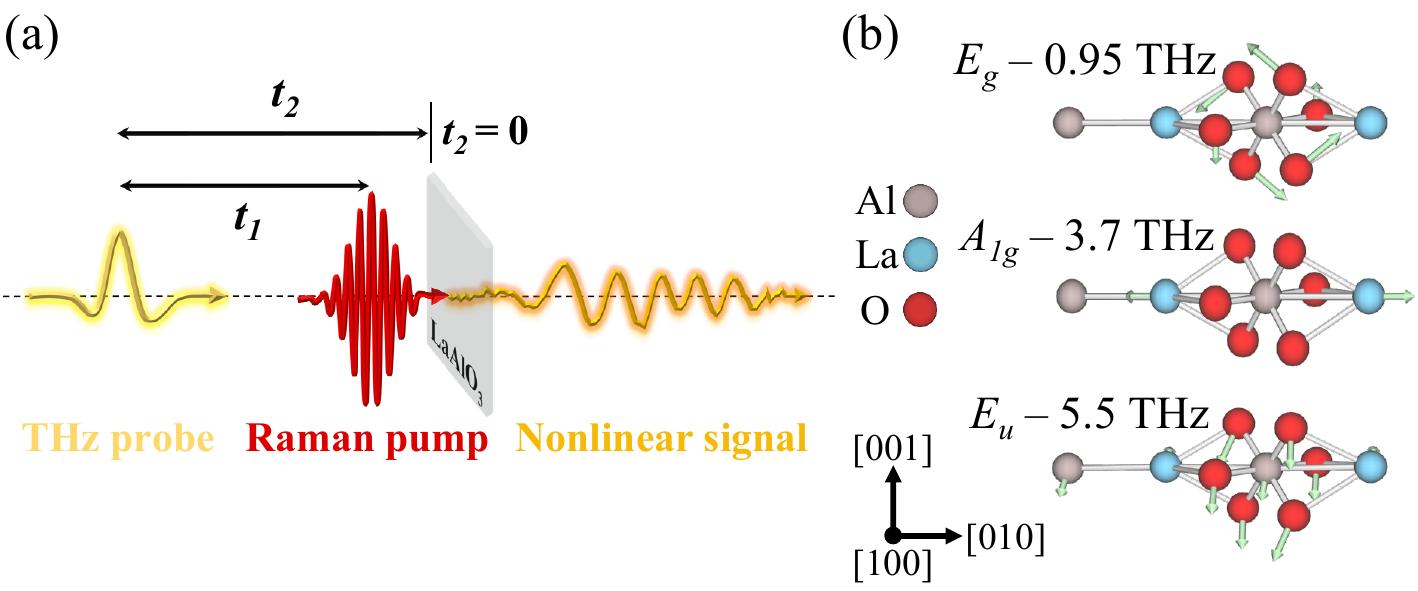}
		\caption{\label{Fig1} (a) Schematic of the experimental geometry for hybrid 2D Raman-THz spectroscopy with the Raman-THz-THz (RTT) pulse sequence. \textit{t}$_1$ is the delay between the impulsive Raman and THz interactions. \textit{t}$_2$ is the delay between the THz interaction and the emitted nonlinear THz field (electro-optic delay time). (b) Calculated phonon displacement patterns for the $E_g$, $A_{1g}$, and $E_{u}$ modes in rhombohedral LAO crystal viewed from [100] direction (adapted from \cite{petretto2018high,Miranda_phononwebsite,Jain2013}). For clarity, only a single oxygen octahedron is shown. } 
\end{figure}

\textit{2D Raman-THz spectroscopy of LAO}— The hybrid 2D-RTT spectroscopy setup is schematized in Fig.\ \ref{Fig1}(a). In this geometry, a Raman pulse is first directed onto the LAO sample to coherently excite Raman-active phonon modes, evolving during the excitation delay $t_1$. Following a second interaction with a time-delayed THz pulse, IR-active phonons can be driven by the coherent modes through anharmonic coupling, resulting in the emission of nonlinear THz fields that are measured over the detection delay $t_2$ via electro-optic sampling (EOS). Anharmonic couplings manifest as off-diagonal features, the cross-peaks, in the 2D-RTT spectrum obtained from a 2D Fourier transformation (FFT) of the time-domain data \cite{Sidler_2019, Sidler_2020}. The phonon spectrum of LAO comprises two soft Raman-active modes that can be efficiently driven by our optical pump at room temperature: the non-fully-symmetric $E_g$ mode at 0.95 THz and the fully symmetric $A_{1g}$ mode at 3.7 THz \cite{Scott_1969, liu1995impulsive, Abrashev_1999, Hayward_2005}. In addition, two IR-active phonons are present: the $A_{2u}$ at 5 THz and the $E_u$ mode at 5.5 THz \cite{lloyd2014modifying}. In Fig.\ \ref{Fig1}(b) we show the displacement patterns of the $E_g$, $A_{1g}$, and $E_u$ modes in the rhombohedral phase.

\begin{figure*}[t]
		\includegraphics[width=0.98\textwidth]{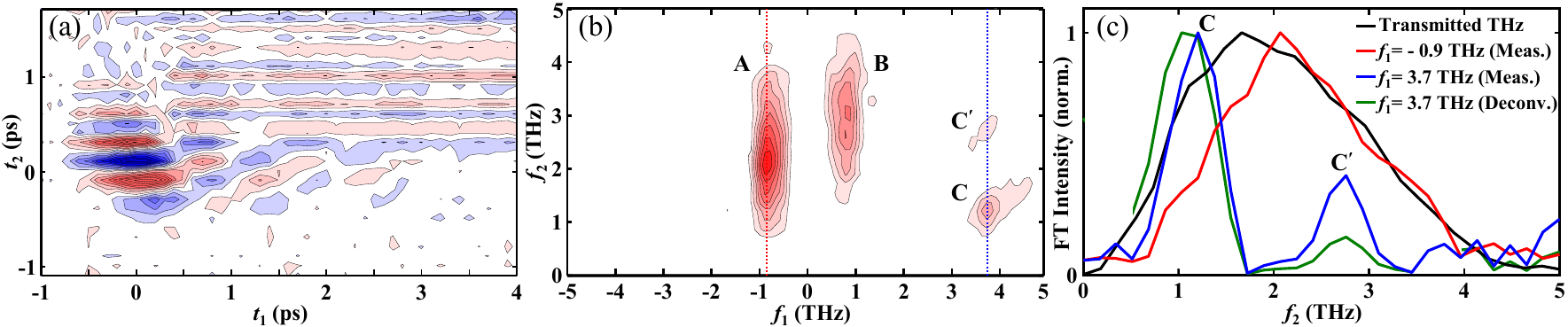}
		\caption{\label{Fig2}(a) Experimental 2D Raman-THz time-domain data measured using the RTT pulse sequence for the LAO crystal at room temperature. (b) Experimental 2D Raman-THz spectrum of LAO. The absolute value of the 2D Fourier transformation for the time-domain data shown in panel (a), obtained after frequency filtering, is plotted. (c) Comparison of measured 1D vertical cuts along the $f_2$ axis for $f_1=-0.9\text{ THz}$ (peak A, red) and $f_1=3.7\text{ THz}$ (peaks C and C$'$, blue) with the transmitted THz spectrum through the LAO crystal (black). Deconvoluted vertical cut for $f_1=3.7\text{ THz}$ is shown in green. The regions where the IRF is too small, and thus, deconvolution is unreliable, are not shown.}
\end{figure*}

The time-domain 2D-RTT response for a room-temperature 50~$\mu $m-thick LAO (100) crystal is reported in Fig.\ \ref{Fig2}(a). The measured response consists of two main components: one featuring strong alternating positive and negative lobes observed within the pulse-overlap region ($-0.7\,\text{ps} < t_1 < 0.4\,\text{ps}$), and the other exhibiting coherent phonon dynamics that extend well beyond this region along the \textit{t}$_1$ axis while remaining short-lived along \textit{t}$_2$. The signal in the pulse-overlap region is dominated by the convolution of the sample's instantaneous response with the THz and Raman pulses. 

To focus only on the coherent phonon dynamics and isolate their oscillatory components, we subtracted a single-exponential fit along each \textit{t}$_1$ cut and applied bandpass filters (from 0.5 to 4.8 THz) along the Raman (\textit{t}$_1$) and the THz (\textit{t}$_2$) axes before performing the 2D FFT. These steps are especially crucial for removing the zero-frequency contribution and suppressing a low-frequency feature at $\sim$0.3 THz, attributed to a subharmonic caused by parametric phonon excitation in LAO \cite{basini2024terahertz}, which would otherwise distort the resulting 2D spectrum. The absolute value of the FFT after the frequency filtering is shown in Fig.\ \ref{Fig2}(b).
Four main features are clearly visible in the 2D spectrum: a strong peak in the rephasing quadrant ($f_1 < 0$ and $f_2 > 0$), labeled A, together with three weaker peaks in the non-rephasing quadrant ($f_1 > 0$ and $f_2 > 0$), labeled B, C, and C$'$. The peaks are narrow along the $f_1$ frequency axis, with frequencies of $f_1=\pm0.9\text{ THz}$ and $f_1=3.7\text{ THz}$, in good agreement with the $E_g$ and $A_{1g}$ phonon frequencies of LAO, respectively.


The key observation is that, while peaks A and B are significantly broader along the $f_2$ axis, peaks C and C$'$ exhibit a round narrow lineshape.
The comparison between 1D vertical cuts along $f_2$ axis for $f_1=-0.9\text{ THz}$ (peak A) and $f_1=3.7\text{ THz}$ (peaks C and C$'$) with the transmitted THz spectrum through the LAO crystal is shown in Fig.\ \ref{Fig2}(c). For $f_1=-0.9\text{ THz}$, the 1D cut shows a similar spectral shape, with comparable bandwidth and peak position to the transmitted THz spectrum (Fig.\ \ref{Fig2}(c), red and black lines).
However, the
signal observed at $f_1=3.7\text{ THz}$ behaves very differently from peaks A and B and does not follow the transmitted THz spectrum (Fig.\ \ref{Fig2}(c), blue). Instead, it has a significantly narrower bandwidth and is shifted towards lower frequencies, with the main-peak position (peak C) centered at $f_2\sim1.2 \text{ THz}$. 
To extract the accurate positions of cross-peaks C and C$^\prime$, we divide the measured response by the instrument response function (IRF) in frequency domain \cite{Ciardi_2019, Shalit_2021, Mousavi_2022, Mousavi_2024}. Details are provided in the End Matter. 
After deconvolution, cross-peak C is found at $f_2\sim1$ THz (Fig.\ \ref{Fig2}(c), green), consistent with the characteristic frequency of the $E_g$ phonon. Frequency position of peak C$^\prime$ remains unchanged, although its relative magnitude with respect to peak C is reduced. 

\textit{Theoretical modeling of the 2D maps}— 
To model the effects of nonlinear phonon-phonon interactions on the 2D spectral maps,
we introduce an anharmonic coupling potential between an IR-active phonon $\mu$ and a Raman-active phonon $\nu$ (Cartesian indices are suppressed for compactness),
\begin{align}\label{eqs:anh_pot}
    V_\text{anh}= g_{\mu\nu} Q_{\text{IR},\mu}^2Q_{\text{R},\nu} + b_{\mu\nu} Q_{\text{IR},\mu}Q_{\text{R},\nu}\text E.
\end{align}
The first term represents a third-order phonon interaction, mediated by the coupling constant $g_{\mu\nu}$, and constitutes a form of mechanical anharmonicity \cite{Martin1974, Juraschek_2018,giorgianni25}. It describes the rectification potential exerted by driven modes on an otherwise undriven phonon. 
The second term corresponds to electrical anharmonicity \cite{khalsa2021,hamm2023}, i.e., a mixed photon-phonon nonlinear interaction. The coupling constant $b_{\mu\nu}\sim dZ_\mu/dQ_{\text{R},\nu}$ encodes the modulation of the Born effective charge $Z_\mu$ of the $\mu$-th mode induced by the $\nu$-th Raman phonon \cite{blank2023two}. Through this mechanism, undriven modes can be excited by field-driven modulation of the lattice polarizability.
  
\begin{figure*}[t]
    \includegraphics[width=0.9\textwidth]{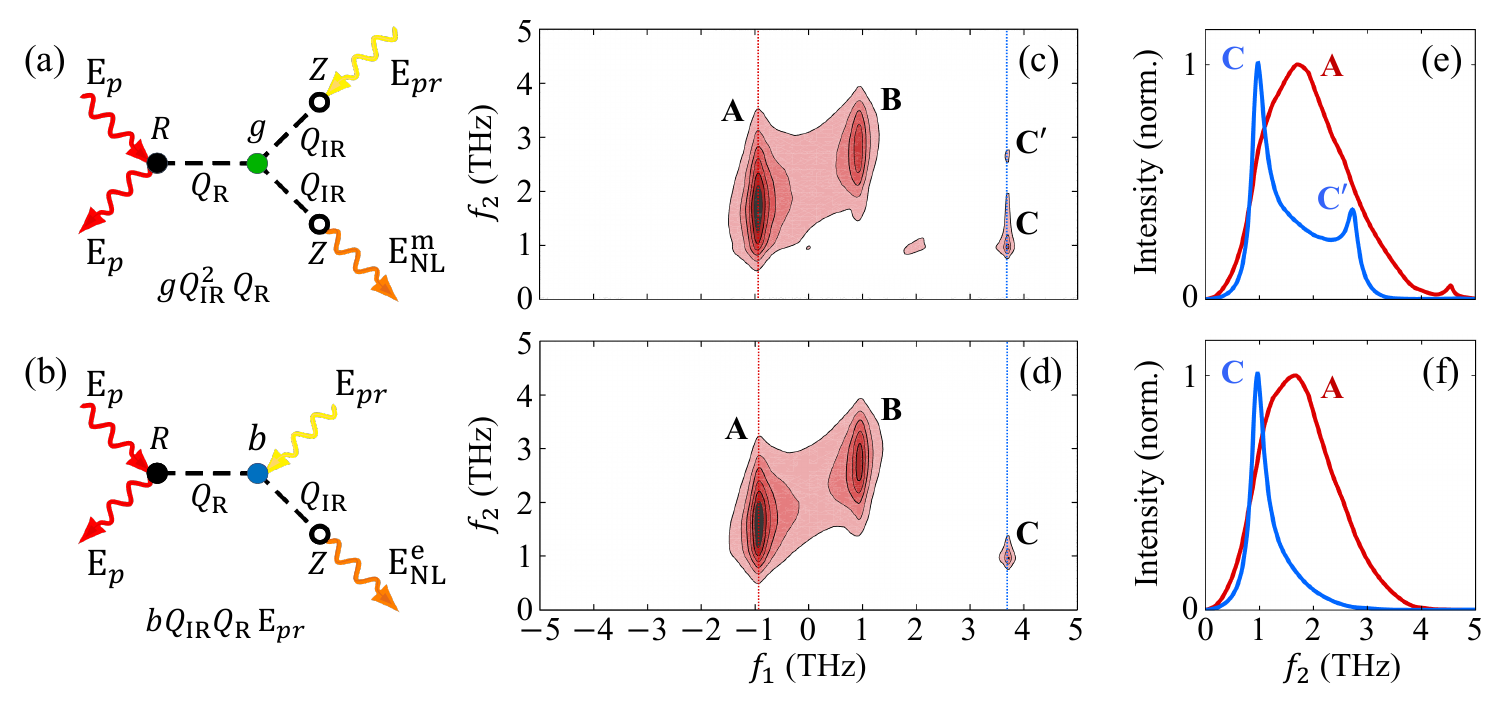}
	\caption{\label{Fig3} (a,b) Diagrammatic representations of the third-order processes involving (a) three-phonon mechanical anharmonicity and (b) electrical anharmonicity. Wavy lines represent photon fields, dashed lines represent phonon fields. 
    (c,d) Absolute value of the simulated 2D response of LAO given by $Q_{\text{IR},E_u}^2Q_{\text{R},E_g}$ and $Q_{\text{IR},E_g}^2Q_{\text{R},A_{1g}}$ as for Eq.\ \eqref{eqs:2Dmech} (panel (c)), and by $Q_{\text{IR},E_u}Q_{\text{R},E_g}\text E_{pr}$ and $Q_{\text{IR},E_g}Q_{\text{R},A_{1g}}\text E_{pr}$ as for Eq.\ \eqref{eqs:2Delec} (panel (d)). Damping parameters are set as $\gamma_{E_g}/2\pi=0.10\text{ THz}$, $\gamma_{A_{1g}}/2\pi=0.05\text{ THz}$ and $\gamma_{E_u}/2\pi=0.05\text{ THz}$. Relative intensity between peaks A and C is tuned to match the experiment. (e,f) Corresponding vertical cuts along the $f_2$ axis for $f_1=-0.95\text{ THz}$ (red) and $f_1=3.7\text{ THz}$ (blue).}
\end{figure*}

Since the measured THz radiation arises from the dynamics of the IR-active mode $Q_{\text{IR},\mu}$, we focus on its response in the presence of the anharmonic couplings introduced in Eq.\ \eqref{eqs:anh_pot}.
The resulting equations of motion can be solved perturbatively.
To leading order in $g_{\mu\nu}$ one obtains the third-order process depicted in Fig.\ \ref{Fig3}(a), in which the IR-active mode is driven by two coherently excited phonons, one Raman-active and one IR-active, through mechanical anharmonicity. This interaction constitutes the reciprocal process of IRS, in which resonantly excited IR-active phonons drive a Raman mode through three-phonon scattering \cite{Maradudin_1970, Wallis_1971, Humphreys_1972, forst_2011, Subedi_2014, von_2018}.
Analogously, to leading order in $b_{\mu\nu}$ one finds the third-order process shown in Fig.\ \ref{Fig3}(b). Here the linear response of the IR-active mode to the probe field is modulated by the coherent excitation of the Raman phonon via electrical anharmonicity. This interaction is the reciprocal of IRRS, in which one resonantly driven IR-active phonon modulates the lattice polarizability and excites Raman modes \cite{khalsa2021,blank2023two}. 

These two excitation pathways result in distinct 2D maps. After introducing the excitation and detection time delays \cite{blank2023two,fiore2D2025,proietto2D2026}, the nonlinear signal in the frequency domain generated by mechanical anharmonicity reads
\begin{align}\label{eqs:2Dmech}
    \text E_\text{NL}^\text{m}(f_1,&f_2)\propto Z_\mu^2R_\nu g_{\mu\nu}D_{\nu}(f_1)D_{\mu}(f_2-f_1)D_{\mu}(f_2)\nonumber\\
    &\times\text E_{pr}(f_2-f_1)\int df^\prime \,\text E_p(f^\prime) \text E_p(f_1-f^\prime),
\end{align}
while that generated by electrical anharmonicity reads
\begin{align}\label{eqs:2Delec}
    \text E_\text{NL}^\text{e}(f_1,&f_2)\propto Z_\mu R_\nu b_{\mu\nu}D_{\nu}(f_1)D_{\mu}(f_2)\nonumber\\
    &\times\text E_{pr}(f_2-f_1)\int df^\prime \,\text E_p(f^\prime) \text E_p(f_1-f^\prime).
\end{align}
Here, $R_\nu$ is the Raman tensor of the $\nu$ mode, $\text E_{pr}(f)$ and $\text E_{p}(f)$ are the probe and pump fields respectively, while $D_\alpha(f)=[f^2-f_\alpha^2+i\gamma_\alpha f/\pi]^{-1}$ is the propagator of the $\alpha$-th phonon ($\alpha=\mu,\nu$), with $f_\alpha$ its characteristic frequency and $\gamma_\alpha$ its damping rate. 
As discussed in the End Matter, the propagators translate into vertical ($D_\nu(f_1)$), horizontal ($D_\mu(f_2)$), or diagonal ($D_\mu(f_2-f_1)$) stripes in the 2D maps, centered at the characteristic phonon frequencies, whose intersections give rise to cross-peaks. Cross-peaks thus appear only when multi-phonon processes are involved, i.e., in the presence of anharmonic couplings. 
In the following we consider the couplings $Q_{\text{IR},E_u}^2Q_{\text{R},E_g}$ and $Q_{\text{IR},E_u}Q_{\text{R},E_g}\text E_{pr}$ between the Raman-active $E_g$ mode ($f_{E_g}=0.95\text{ THz}$) and the IR-active $E_u$ phonon ($f_{E_u}=5.5\text{ THz}$), and the couplings $Q_{\text{IR},E_g}^2Q_{\text{R},A_{1g}}$ and $Q_{\text{IR},E_g}Q_{\text{R},A_{1g}}\text E_{pr}$ of the Raman-active $A_{1g}$ mode ($f_{A_{1g}}=3.7\text{ THz}$) with the $E_g$ phonon. The latter are allowed only if the $E_g$ mode acquires a finite IR activity, as justified in the next Section.

To model the experimental response, we use the experimental pump and probe pulses and account for the filtering of both the sample and the THz detection crystal along $f_2$. 
In Figs.\ \ref{Fig3}(c,d), we show the simulated 2D maps for the excitation pathways described above. 
Most of the frequency structure associated with processes mixing $Q_{\text{IR},E_u}$ and $Q_{\text R,E_g}$ is filtered out at high $f_2$. Within the bandwidth of our THz probe, only the vertical stripes given by $D_{E_g}(f_1)$ remain visible (see Fig.\ \ref{Fig9} in the Supplemental Material for the frequency structure of the propagators only). These give rise to the broad peaks A and B, which thus correspond to a direct observation of the Raman excitation of $Q_{\text{R},E_g}$ occurring in both excitation pathways. 
Vertical cuts along the $f_2$ axis for $f_1=-0.95 \text{ THz}$ (peak A) are shown in Figs.\ \ref{Fig3}(e,f) (red lines), exhibiting a spectral shape comparable to that of the probe, as observed in Fig.\ \ref{Fig2}(c). This behavior comes from the convolution of the phonon propagators, approximately constant in $f_2$ far from $f_{E_u}$, and the probe $\text{E}_{pr}(f_2-f_1)$.
Cross-peak C arises from anharmonic processes mixing $Q_{\text{IR},E_g}$ and $Q_{\text{R},A_{1g}}$, while the secondary cross-peak C$^\prime$ originates from the three-phonon process between these modes, due to the presence of the additional propagator $D_{E_g}(f_2-f_1)$. Vertical cuts along $f_2$ for $f_1=3.7\text{ THz}$ are shown in Figs.\ \ref{Fig3}(e,f) (blue lines). 
It has been argued in Ref.\ \cite{hamm2023} that mechanical anharmonicity vanishes for perfectly delocalized phonons in the low-excitation limit, as the atomic displacements associated with the phonon mode tend to zero. The observation of cross-peak C$^\prime$ therefore indicates a finite degree of phonon localization, and that both electrical and mechanical anharmonicities contribute to the 2D signal.
Notice that the three-phonon process also predicts the appearance of an additional peak at $(2f_{E_g},f_{E_g})$, which is not observed in the experimental spectrum. This discrepancy is likely due to the peak being significantly suppressed by the reduced IRF intensity at that point, compared with the position of the C$^\prime$ cross-peak. 

Processes that do not involve anharmonic couplings between the $A_{1g}$ and the $E_g$ modes cannot account for the position and width of the cross-peaks C and C$^\prime$. In the Supplemental Material we present the calculated 2D map for the Raman-like (two-photon) excitation and detection of the $A_{1g}$ phonon, as it occurs in impulsive stimulated Raman scattering, which fails to reproduce the experimental 2D spectrum.

\textit{IR activity on LAO domain walls}—In centrosymmetric LAO, third-order optical processes mediated by anharmonic couplings between the $A_{1g}$ mode and the nominally Raman-active $E_g$ phonon \cite{hayward2002order,Hayward_2005,Abrashev_1999,basini2024terahertz,liu1995impulsive} are forbidden by symmetry. 
Nonetheless, the presence of cross-peaks in the 2D spectrum suggests that the $E_g$ mode acquires a finite IR activity.
This condition can be realized if inversion symmetry is locally broken by structural DWs, where the additional lattice distortions can induce a finite dipolar activity of otherwise Raman-active phonons.
In its rhombohedral phase, LAO displays a tilting pattern of AlO$_6$ octahedra denoted as $a^-a^-a^-$ in Glazer notation \cite{Glazer1975}, corresponding to antiphase rotations around one of the space diagonal (pseudo)cubic (PC) axes of the perovskite structure. Static rotational momenta define the tilting order parameter $\boldsymbol{\Phi}$ \cite{litvin2011,iniguez2013,barone2014}, an axial vector whose direction indicates the rotation axis of the O$_6$ octahedra and whose magnitude gives the rotation angle. DWs separate two domains with the same tilting pattern and rotation angle $(4.7^\circ)$, but different rotation axes. 

\begin{figure}[t]
		\includegraphics[width=0.48\textwidth]{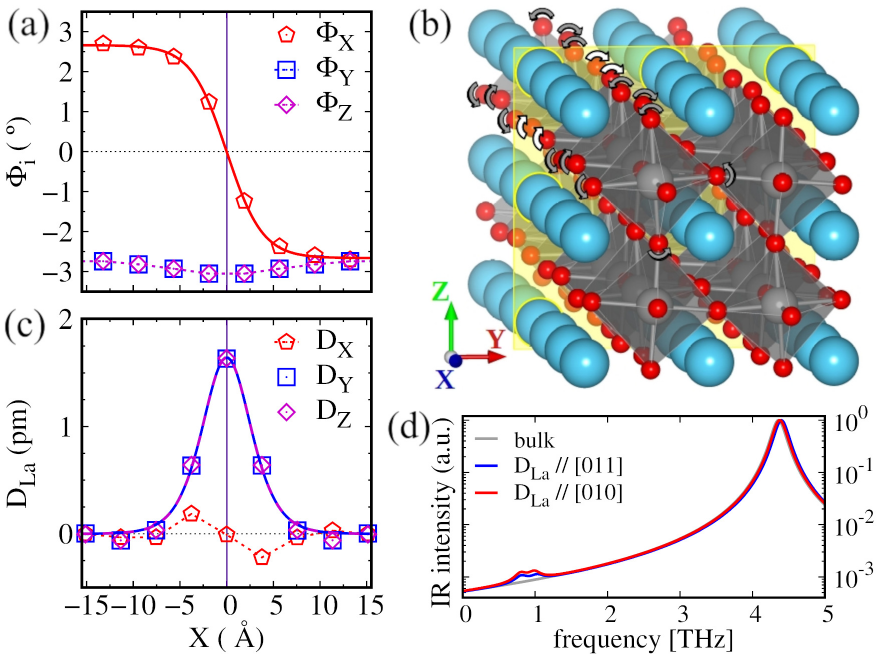}
        \caption{\label{figure_dft}(a) Evolution across a DW (at X=0 \AA) of the tilting order parameter $\boldsymbol{\Phi}$, describing antiphase rotations of AlO$_6$ octahedra around the three PC directions X, Y and Z. The X axis is perpendicular to the DW, across which the antiphase rotation $\Phi_X$ changes sign.
        The red line shows the $\Phi_X$ profile described by $\Phi_0\tanh(X/W)$, with a domain-wall width of $2W=7.7$~\AA. (b) Sketch of the structural changes across the DW as obtained by DFT calculations. La atoms are contained within the DW plane. Gray arrows depict the antiphase rotation of O$_6$ octahedra around PC direction X, while white arrows highlight the in-phase rotation across the wall. (c) Corresponding offcentering of La sublattice, occurring within the twin-wall plane along the $[011]$ direction and displaying the characteristic $\sech(X/W)^{-2}$ (solid lines) profile of a secondary order parameter coupled through a biquadratic coupling to the primary one \cite{Salje_book}. (d) Simulated vibrational IR spectrum of bulk rhombohedral LAO (gray), and bulk LAO with the La sublattice offcentered along $[011]$ (blue) or $[010]$ (red). Because of the reduced symmetry at the DW, the twofold degeneracy of the $E_g$ mode is lifted, and two close peaks are observed. The spectra are normalized by the dominating $A_{2u}$ mode at the frequency $\sim 4.4$ THz obtained with DFT.}
\end{figure}

To address whether the $E_g$ mode can become IR-active, we perform first-principles DFT calculations on two distinct LAO domain-wall configurations, following the approach of Ref.\ \cite{barone2014}. Details of the calculations are provided in the Supplemental Material. We consider a 320-atom supercell hosting two DWs perpendicular to the PC direction X.
On the first configuration considered, the rotation axis changes from [$\bar{1}$11] to [111] across the DW, as shown in Fig.\ \ref{figure_dft}(a). This implies an in-phase rotation of O$_6$ octahedra across the interface, locally realizing a $c^+a^-a^-$ pattern, illustrated in Fig.\ \ref{figure_dft}(b). The latter modifies the first oxygen-coordination shell of the La cation, inducing its off-centering along the [011] direction \cite{iniguez2013}.
This can be quantified by defining a secondary order parameter $\textbf{D}$ from the local displacement of La atoms with respect to the center of mass of the surrounding  dodecahedral coordination cage LaO$_{12}$. The evolution of $\textbf{D}$ across the twin domains is shown in Fig.\ \ref{figure_dft}(c), revealing the polar character of the DW.
In the second configuration, discussed in the Supplemental Material, the rotation of the tilting axis from $[111]$ to $[\bar{1}1\bar{1}]$ across the DW would instead realize a local $b^+a^-c^+$ tilt pattern and an A-site offcentering parallel to $[010]$ \cite{iniguez2013}.

Even though the induced displacement is on the order of few pm, the combination of anomalously large effective charges of the La cations ($\sim 4.4 \vert e\vert$) and O anions ($\sim -2.5 \vert e\vert$), together with local inversion-symmetry breaking, can support the activation of a finite IR response of the $E_g$ phonon mode, which is associated with the tilting order parameter. 
Due to the high computational cost of directly simulating the IR vibrational spectra of the supercell, we consider a bulk rhombohedral LAO in which the La sublattice is artificially offcentered by 2.5 pm along [011] or [010]. The resulting IR intensities, shown in \ref{figure_dft}(d) and compared with that of bulk LAO, reveal the emergence of a small but finite IR activity of the $E_g$ phonons.  This prediction is further supported by THz emission measurements, reported in the Supplemental Material.

\textit{Conclusion}—We have presented the 2D Raman-THz-THz spectrum of LAO, providing clear experimental evidence of anharmonic couplings between the low-frequency $A_{1g}$ and $E_g$ phonons, revealed by the localized cross-peak features. Numerical simulations show that these signatures arise from a finite IR activity of the $E_g$ mode. Density functional theory calculations further indicate that this IR activation originates from local inversion-symmetry breaking at ferroelastic domain walls of the crystal. 
Our results shed new light on the role of twin domains in the complex lattice dynamics of LAO and related ferroelastic materials, and highlight the potential of 2D-RTT spectroscopy to reveal anharmonic coupling channels associated with subtle symmetry breaking.

\textit{Acknowledgments}—M.U. acknowledges financial support from the French Agence Nationale de la Recherche (ANR) Grant No.\ ANR-23-CE30-0030 (SUPER2DTMD). M.B. acknowledges support from the SNSF (Ambizione project, PZ00P2 216089). P.B. acknowledges the CINECA award under the ISCRA initiative Grant No. HP10CM2RK0 for the availability of high-performance
computing resources and support. N.S. acknowledges financial support from Sapienza University of Rome under the project Ateneo (AR125199BDE01384) and from the Italian MIUR under the project PRIN2022-CoInEx (2022WS9MS4).

\textit{Data availability}—The data that support the findings of this study are not publicly available. The data are available from the authors upon reasonable request.

\vspace{0.5cm}

\bibliography{apstemplate}
\newpage
\section*{End Matter}
\begin{figure}[h]
		\includegraphics[width=0.4\textwidth]{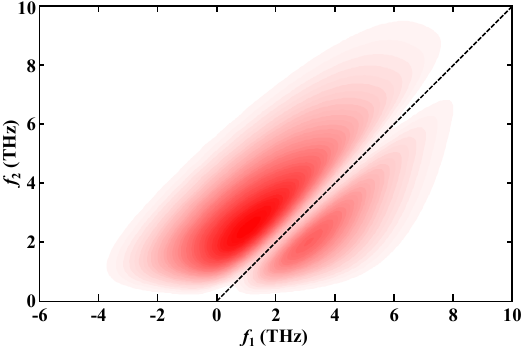}
		\caption{\label{Fig8} Absolute value of the IRF in the frequency domain for the 2D-RTT pulse sequence.}
\end{figure}

\textit{Instrument Response Function}—The absolute value of the instrument response function in the frequency domain for the 2D-RTT pulse sequence is shown in Fig.\ \ref{Fig8} (see Refs.\ \cite{Ciardi_2019} and \cite{Sidler_2019} for details). 
The convolution of the sample response with the IRF affects the frequency position, shape, and amplitude of the observed cross-peaks in the 2D spectrum, thus requiring proper deconvolution.

\begin{figure}[t]
		\includegraphics[width=0.4\textwidth]{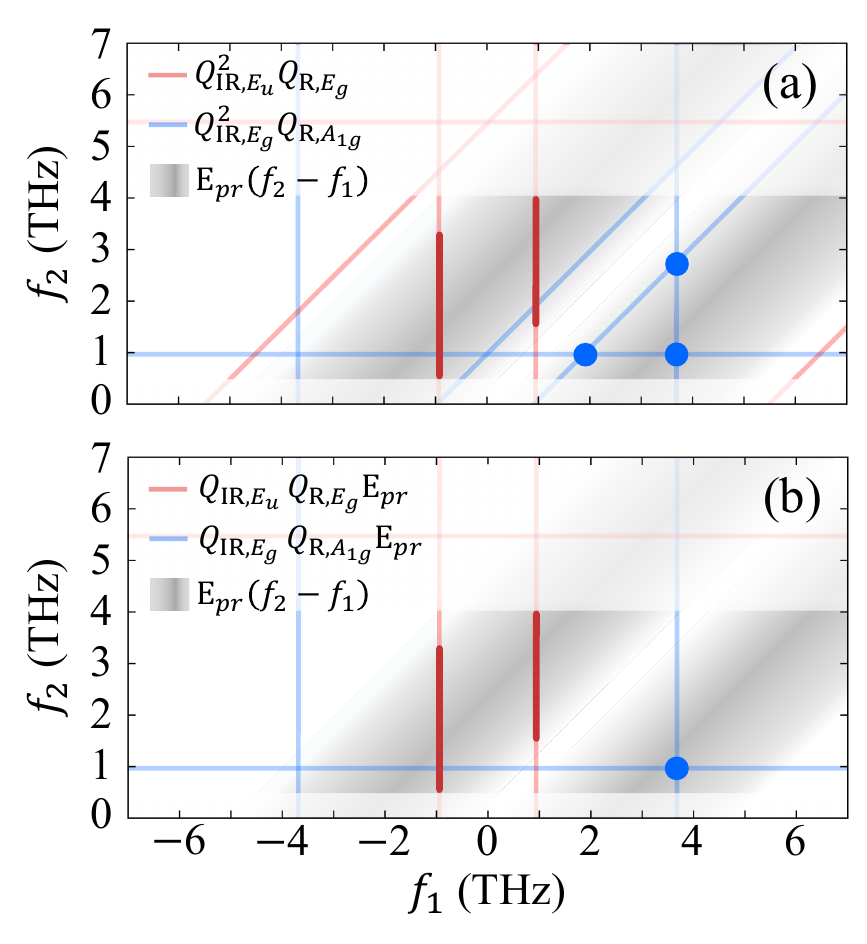}
		\caption{\label{Fig_stripes} (a) Sketch of the contributions to the 2D response $\text E^\text{m}_\text{NL}(f_1,f_2)$ as in Eq.\ \eqref{eqs:2Dmech}. Stripes given by the phonon propagators are shown for the $Q_{\text{IR},E_u}^2Q_{\text{R},E_g}$ process (red lines) and for the $Q_{\text{IR},E_g}^2Q_{\text{R},A_{1g}}$ process (blue lines). (b) Sketch of the contributions to the 2D response $\text E^\text{e}_\text{NL}(f_1,f_2)$ as in Eq.\ \eqref{eqs:2Delec}. Stripes given by the phonon propagators are shown for the $Q_{\text{IR},E_u}Q_{\text{R},E_g}\text E_{pr}$ process (red lines) and for the $Q_{\text{IR},E_g}Q_{\text{R},A_{1g}}\text E_{pr}$ process (blue lines). In both panels, the modulation given by $\text E_{pr}(f_2-f_1)$ is sketched in gray for the experimental THz probe. Dark-red lines and the blue dots highlight the regions of maximum overlap between the propagators and the probe field. White bands display regions that are filtered out by the sample and by the THz detection crystals.}
\end{figure}

\textit{Contributions to the theoretical 2D maps}—In Fig.\ \ref{Fig_stripes} we show a sketch of the different terms that compose Eqs.\ \eqref{eqs:2Dmech} (panel a) and \eqref{eqs:2Delec} (panel b). In particular, we show the contribution of the couplings $Q_{\text{IR},E_u}^2Q_{\text{R},E_g}$ and $Q_{\text{IR},E_u}Q_{\text{R},E_g}\text E_{pr}$ (red lines), and that of the couplings $Q_{\text{IR},E_g}^2Q_{\text{R},A_{1g}}$ and $Q_{\text{IR},E_g}Q_{\text{R},A_{1g}}\text E_{pr}$ (blue lines). These are the anharmonic processes simulated in Fig.\ \ref{Fig3} of the main text. 

The contribution of the electric fields $\text E^3(f_1,f_2)=\text E_{pr}(f_2-f_1)\int df^\prime \,\text E_p(f^\prime) \text E_p(f_1-f^\prime)$ does not depend on the combination of different phonons excited, but rather it stems from the delay scheme of the pulses. In particular, one can directly visualize the modulation of the probe field $\text E_{pr}(f_2-f_1)$ (gray diagonal stripes for the experimental probe).
On the other hand, the propagators $D_\alpha(f)$ encode the contribution of the phonons, giving rise to two stripes each, peaked at $f=\pm f_\alpha$. In particular, $D_\nu(f_1)$ gives vertical stripes at the frequency of the Raman phonon excited by the pump, while $D_\mu(f_2-f_1)$ and $D_\mu(f_2)$ give respectively diagonal and horizontal stripes at the frequency of the IR phonon. 
Peaks are expected, in the 2D spectrum, where the overlap between the phonon propagators and the probe spectrum is maximized. In Fig.\ \ref{Fig9} of the Supplemental Material we show the spectrum obtained with a constant $\text{E}_{pr}(f)=\text{E}_{pr}$, in which all expected cross-peaks given by the anharmonic processes are clearly visible. 
Additionally, the sample and the THz detection crystal filter the nonlinear response along $f_2$, as sketched by the white horizontal bands. The areas of Fig.\ \ref{Fig_stripes} marked by dark-red lines or dark-blue dots correspond to regions of maximum overlap between propagators and probe field that fall within the detection window, which then result in the 2D spectra reported in Figs.\ \ref{Fig3}(c,d).
\clearpage
\section{Supplemental Material}

\section{2D Raman-THz spectrometer}

The two-dimensional time-domain signals were obtained using a custom-built 2D Raman–THz spectrometer that employs rapid-scan multichannel detection combined with a small-bias electro-optic sampling, as described in detail in \cite{Shukla_2025}.
In brief, the output of an amplified 100 kHz Yb-laser system, operating at a center wavelength of 1030~nm with a pulse duration of 130~fs, is split into two branches: one for Raman excitation and one for THz generation and detection. Broadband pulses covering frequencies of up to 5~THz are generated through optical rectification in a 200~$\mu$m thick (110) GaP crystal. The generated THz field is then focused onto the sample collinearly with the 20~$\mu$J Raman excitation pulse and then directed to the detection crystal [identical 200~$\mu$m-thick (110) GaP crystal] using two custom-made aluminum elliptical mirrors (2f = 83~mm).  The temporal delay (\textit{t}$_1$) between the Raman excitation and the generated THz field is controlled by a step-scanning delay motor (M-405.DG, Physik Instrumente). The generation and Raman beams were modulated using a pair of mechanical choppers (310CD, Scitec Instruments), operating at 25 and 50 kHz, respectively, to isolate the third-order nonlinear response according to the signal calculation scheme described in \cite{Mousavi_2024}. Delay-zero with regard to \textit{t}$_1$ and \textit{t}$_2$ is defined as the peak of the instrument response function.
To facilitate multichannel detection, the gated beam is expanded by a factor of 16 and then passed through two transparent  15 × 15 mm² BK7 glass echelon masks with variable thickness, forming a six-by-six array of beamlets that spans a total delay range of $\sim$1.4~ps before reaching the detection crystal \cite{Duchi_2021}. Subsequently, the gated beam is imaged onto a pair of custom-built photodetectors \cite{Farrell_2020} composed of 32 silicon photodiodes arranged spatially to match the shape of the beamlets. The change in polarization across all pixels of the gated beam, associated with the strength of the THz field, was measured using a small-bias electro-optic sampling scheme \cite{Mousavi_2024}. By combining this detection scheme with the multichannel approach, B-matrix referencing, as described in Ref.~\cite{Feng_2019}, can be applied, which both significantly enhances the spectrometer’s signal-to-noise ratio and increases its tolerance to misalignment of the 32 individual beamlets across the two cameras. Finally, to compensate for nonuniformities between individual beamlets such as variations in delay, noise, and intensity arising from the nonuniform properties of the expanded beam, the gate delay time (\textit{t}$_2$) was continuously varied using an electromagnetically driven stage (V-408, Physik Instrumente) at a rate of 4~Hz. This method enabled the rapid acquisition of 32 time-shifted THz profiles, which were then combined during post-processing to generate an extended and uniform THz profile.


%


\section{Raman pulse fluence dependence}
\begin{figure}[t]
		\includegraphics[width=0.35\textwidth]{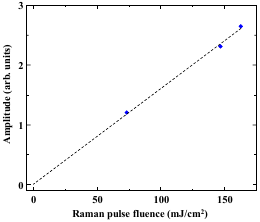}
		\caption{\label{Fig6} Raman pulse fluence dependence of the (3.7 THz, 1 THz) cross-peak intensity. Signal shows a linear dependence on the Raman pump fluence ($E^2$ process).}
\end{figure}

The 2D Raman–THz signal involves two field interactions with the Raman pulse and therefore is expected to depend linearly on the Raman pulse fluence \cite{Savolainen_2013}. Figure~\ref{Fig6} shows the fluence dependence of the observed cross-peak at (3.7 THz, 1 THz). The experimental data, plotted as solid blue circles, exhibit a clear linear correlation with the Raman pulse fluence, consistent with the behavior expected for a 2D Raman–THz response. This linear dependence is confirmed by a fitted linear curve (black dashed line).
\section{Details of the theoretical analysis}
\subsection{Perturbative solution of equations of motion}
The anharmonic scattering processes between an IR-active phonon $\mu$ and a Raman-active phonon $\nu$ are described by the potential Eq.\ \eqref{eqs:anh_pot}, that generates anharmonic driving terms that couple the two modes in the equations of motion.
In frequency domain, these read explicitly
\begin{align}
    D_\mu^{-1}(\omega)Q_{\text{IR},\mu}(\omega)+b_{\mu\nu}\int d\omega^\prime\, \text E(\omega-\omega^\prime)Q_{\text{R},\nu}(\omega^\prime)&\nonumber\\
    +2g_{\mu\nu}\int d\omega^\prime Q_{\text{IR},\mu}(\omega-\omega^\prime)Q_{\text{R},\nu}(\omega^\prime)=Z_\mu\text{E}(\omega), \label{eqs:motion_frequency1}&\\
    D_\nu^{-1}(\omega)Q_{\text{R},\nu}(\omega)+b_{\mu\nu}\int d\omega^\prime\, \text E(\omega-\omega^\prime)Q_{\text{IR},\mu}(\omega^\prime)&\nonumber\\
    +g_{\mu\nu}\int d\omega^\prime Q_{\text{IR},\mu}(\omega-\omega^\prime)Q_{\text{IR},\mu}(\omega^\prime)=R_\nu\text{E}^2(\omega)&,\label{eqs:motion_frequency2}
\end{align}
where $\omega$ is the frequency associated with the time $t$, $D_\mu(\omega)=[\omega^2-\omega_\mu^2+2i\gamma_\mu\omega]^{-1}$ and $D_\nu(\omega)=[\omega^2-\omega_\nu^2+2i\gamma_\nu\omega]^{-1}$ are the propagators of the IR-active and Raman-active phonons respectively, $R_\nu$ is the Raman tensor associated with the Raman-active mode, and $\text E^2(\omega)=\int d\omega^{\prime\prime}\text E(\omega^{\prime\prime})\text E(\omega-\omega^{\prime\prime})$. 
To maintain a compact notation, we have kept the dependence on $t_1$ and $t_2$ implicit here.

To gain an intuitive understanding of the 2D spectra resulting from such coupled equations, it is useful to derive an analytical solution of the equations of motion. Thus, we assume small anharmonic coupling constants, and we solve Eqs.\ \eqref{eqs:motion_frequency1} and \eqref{eqs:motion_frequency2} to leading order in $g_{\mu\nu}$ or $b_{\mu\nu}$. At zero-th order one finds
\begin{align}
    &Q_{\text{IR},\mu}^{(0)}(\omega)=\frac{Z_\mu\text{E}(\omega)}{(\omega^2-\omega_\mu^2)+2i\gamma_\mu\omega},\label{eqs:sol_zero1}\\
    &Q_{\text{R},\nu}^{(0)}(\omega)=\frac{R_\nu\text E^2(\omega)}{(\omega^2-\omega_\nu^2)+2i\gamma_\nu\omega}.\label{eqs:sol_zero2}
\end{align}
We can then solve Eq.\ \eqref{eqs:motion_frequency1} for $Q_{\text{IR},\mu}(\omega)$ to first order in $\text g_{\mu\nu}$,
\begin{align}\label{eqs:sol_one}
    Q_{\text{IR},\mu}^\text{m}(\omega)=\frac{-2g_{\mu\nu}\int d\omega^\prime Q_{\text{IR},\mu}^{(0)}(\omega-\omega^\prime)Q_{\text{R},\nu}^{(0)}(\omega^\prime)}{(\omega^2-\omega_\mu^2)+2i\gamma_\mu\omega},
\end{align}
which represents the IR-active phonon driven by the three-phonon anharmonic scattering. Analogously, we can solve Eq.\ \eqref{eqs:motion_frequency1} to leading order in $b_{ij}$,
\begin{align}\label{eqs:sol_oneb}
    Q_{\text{IR},\mu}^\text{e}(\omega)=\frac{- b_{\mu\nu}\int d\omega^\prime \text E(\omega-\omega^\prime)Q_{\text{R},\nu}^{(0)}(\omega^\prime)}{(\omega^2-\omega_\mu^2)+2i\gamma_\mu\omega},
\end{align}
which represents the IR-active phonon excited by the electromagnetic field and modulated by the Raman-active phonon.

\subsection{Phonon mapping in 2D frequency space}
The driven phonon depends on the time delays between the pulses $Q_{\text{IR},\mu}(\omega)=Q_{\text{IR},\mu}(\omega,t_1,t_2)$ through the electric field $\text E(\omega)=\text E(\omega,t_1,t_2)$, that contains both the probe $\text E_{pr}(\omega)$ and the pump $\text E_{p}(\omega)$ fields. Explicitly,
\begin{align}\label{eqs:el_field}
    \text E(\omega,t_1,t_2)=\text E_p(\omega)e^{-i\omega(t_1+t_2)}+\text E_{pr}(\omega)e^{-i\omega t_2}.
\end{align}
In Eqs.\ \eqref{eqs:sol_one} and \eqref{eqs:sol_oneb} we retain only terms that scale as $\text E_p^2\text E_{pr}$, since the 2D-RTT technique eliminates any other component of the response. By taking the Fourier transform $t_1\to\omega_1$ and $t_2\to\omega_2$, and integrating over $\omega$ as prescribed by the detection process, we find 
\begin{widetext}
\begin{align}\label{eqs:2Dphonon}
    Q_{\text{IR},\mu}^\text{m}(\omega_1,\omega_2)=-2Z_\mu R_\nu g_{\mu\nu}&\text E_{pr}(\omega_2-\omega_1)\int d\omega^\prime \text E_p(\omega^\prime)\text E_{p}(\omega_1-\omega^\prime)\nonumber\\
    \times&\big[D_\nu(\omega_1)D_\mu(\omega_2-\omega_1)D_\mu(\omega_2)+D_\nu(\omega_2-\omega_1+\omega^\prime)D_\mu(\omega_1-\omega^\prime)D_\mu(\omega_2)\big],
\end{align}
\end{widetext}
and, analogously,
\begin{widetext}
\begin{align}\label{eqs:2Dphononb}
    Q_{\text{IR},\mu}^\text{e}(\omega_1,\omega_2)=-R_\nu b_{\mu\nu}\text E_{pr}(\omega_2-\omega_1)\int d\omega^\prime \text E_p(\omega^\prime)\text E_{p}(\omega_1-\omega^\prime)\big[D_\nu(\omega_1)D_\mu(\omega_2)+D_\nu(\omega_2-\omega_1+\omega^\prime)D_\mu(\omega_2)\big].
\end{align}
\end{widetext}
Upon closer inspection of these two solutions, one recognizes the second term within the square brackets to be associated with the process in which the Raman-active phonon is driven by the combination of a pump and a probe photon. Since the spectral content of the pump is in the eV range, and that of the probe is of order of THz, such two-photon excitation of the Raman-active mode is off-resonant, making this process subleading. On the other hand, the first is the dominant term, as it is associated with the process in which the Raman-active phonon is driven by two pump photons through a difference-frequency excitation. We thus retain only the latter in the following.

\begin{figure}[t]
		\includegraphics[width=0.49\textwidth]{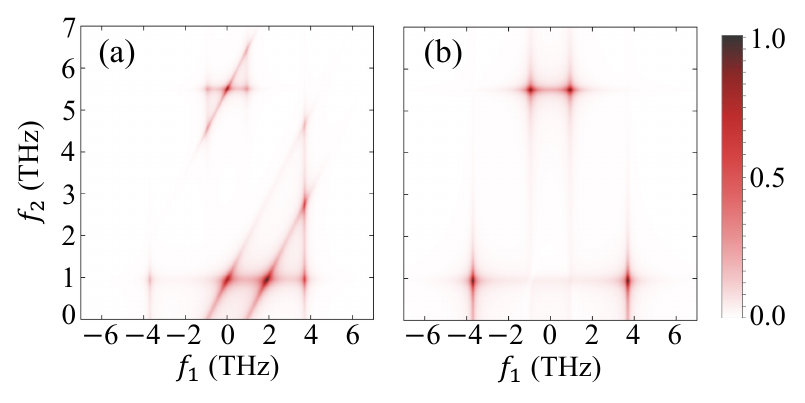}
		\caption{\label{Fig9} (a,b) Absolute value of the simulated 2D response of LAO given by processes with mechanical (a) and with electrical anharmonicity (b), for a constant probe field $\text{E}_{pr}(f)=\text{E}_{pr}$ and without the $f_2$ filtering given by sample and detection crystal. Damping parameters are set as $\gamma_{E_g}/2\pi=0.10\text{ THz}$, $\gamma_{A_{1g}}/2\pi=0.05\text{ THz}$ and $\gamma_{E_u}/2\pi=0.05\text{ THz}$. All anharmonic processes are normalized to their respective maximum value.}
\end{figure}

The emitted nonlinear 2D signal can be connected to the anharmonically-driven phonon as $\text E_\text{NL}(\omega_1,\omega_2)\propto Z_\mu Q_{\text{IR},\mu}(\omega_1,\omega_2)$. By shifting $\omega\to2\pi f$, one then retrieves Eqs.\ \eqref{eqs:2Dmech} and \eqref{eqs:2Delec} of the main text from Eqs.\ \eqref{eqs:2Dphonon} and \eqref{eqs:2Dphononb} respectively. In Fig.\ \ref{Fig9} we show the simulated spectrum obtained from Eqs. \eqref{eqs:2Dmech} and \eqref{eqs:2Delec} for a constant probe field $\text{E}_{pr}(f)=\text{E}_{pr}$, to clearly show the cross-peaks given by the intersections between the phonon propagators. 
To simulate the true experimental response, the nonlinear signal $\text E_\text{NL}(f_1,f_2)$ is multiplied by a transmission factor $\text T(f_2)$, to take into account the filtering performed by the sample and by the THz detection crystal in which the EOS is performed. 
We note that $\text E^\text{m}_\text{NL}(f_1,f_2)$ scales overall as $Z_\mu^2R_\nu g_{\mu\nu}$, while $\text E^\text{e}_\text{NL}(f_1,f_2)$ scales as $Z_\mu R_\nu b_{\mu\nu}$. Explicit calculation of the ratio $g_{\mu\nu}Z_\mu/b_{\mu\nu}$ at the LAO domain walls methods could provide an estimate of the dominant excitation pathway.

\subsection{ISRS-like excitation of the $A_{1g}$ mode}
As discussed in the main text, the observed position and width of cross-peaks C and C$^\prime$ are only understood if an anharmonic interaction between the Raman-active $A_{1g}$ phonon and the IR-active $E_g$ mode is taken into account. 

\begin{figure}[t]
		\includegraphics[width=0.47\textwidth]{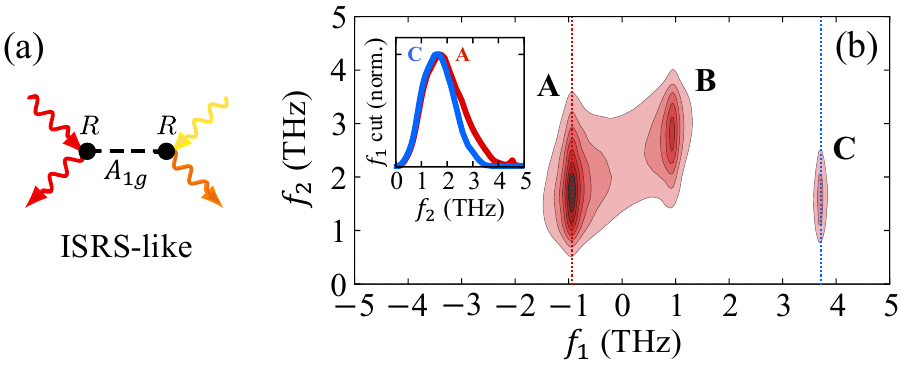}
		\caption{\label{Fig7} (a) Diagrammatic representation of the ISRS-like third-order process involving a two-photon excitation and Raman-like detection of the $A_{1g}$ phonon. (b) Absolute value of the simulated 2D response of LAO. Peaks A and B are given by three-phonon mechanical anharmonicity as for Eq.\ \eqref{eqs:2Dmech}, while peak C is given by the ISRS-like process as for Eq.\ \eqref{eqs:2Disrs}. The inset shows vertical cuts along the $f_2$ axis for $f_1=-0.95\text{ THz}$ (red) and $f_1=3.7\text{ THz}$ (blue). Damping parameters are set as $\gamma_{E_g}/2\pi=0.10\text{ THz}$, $\gamma_{A_{1g}}/2\pi=0.05\text{ THz}$ and $\gamma_{E_u}/2\pi=0.05\text{ THz}$.}
\end{figure}

We here consider a two-photon excitation and Raman-detection process involving the $A_{1g}$ mode only, as depicted in Fig.\ \ref{Fig7}(a). This is the typical excitation pathway of impulsive stimulated Raman scattering (ISRS), although performed with a THz probe. In this case, the equations of motion for the Raman-active mode read
\begin{align}\label{eqs:raman_eqs}
    D_{A_{1g}}^{-1}(\omega)Q_{\text{R},A_{1g}}(\omega)=R_{A_{1g}}\text E^2(\omega),
\end{align}
which can be solved analytically. The nonlinear signal in 2D space can be connected to the driven Raman phonon as $\text E_\text{NL}(\omega_1,\omega_2)\propto R_{A_{1g}}Q_{\text{R},A_{1g}}(\omega_1,\omega_2)\text E(\omega_1,\omega_2)$, reading explicitly ($\omega\to2\pi f)$
\begin{align}\label{eqs:2Disrs}
    \text{E}^\text{ISRS}_\text{NL}(f_1,f_2)\propto R_{A_{1g}}^2D_{A_{1g}}(f_1)\text E^3(f_1,f_2).
\end{align}
In this process the only phonon modulation is given by $D_{A_{1g}}(f_1)$, corresponding to vertical stripes at $f_1=\pm f_{A_{1g}}=\pm 3.7\text{ THz}$. Once convoluted with the probe, these result in a broad feature at $f_1=f_{A_{1g}}$, with spectral shape comparable with the probe spectrum, see Fig.\ \ref{Fig7}(b). This is in contrast with the experimental observation of two narrow and distinct peaks, indicating that the ISRS-like process is subleading for the $A_{1g}$ phonon of LAO.

We note that ISRS-like scattering cannot be at the origin of peaks A and B at $f_{1}=\pm f_{E_g}$. Because the Raman tensor associated with the $E_g$ mode is off-diagonal \cite{liu1995impulsive}, the THz probe would be scattered with its polarization rotated by $90^\circ$. Since our detection scheme measures THz transmission changes in the same polarization direction of the incoming probe, such ISRS-like signals from the $E_g$ mode cannot be measured. Peaks A and B shown in Fig.\ \ref{Fig7}(b) are obtained from Eq.\ \eqref{eqs:2Dmech} of the main text.

\section{Computational details}
\begin{figure}[t]
		\includegraphics[width=0.45\textwidth]{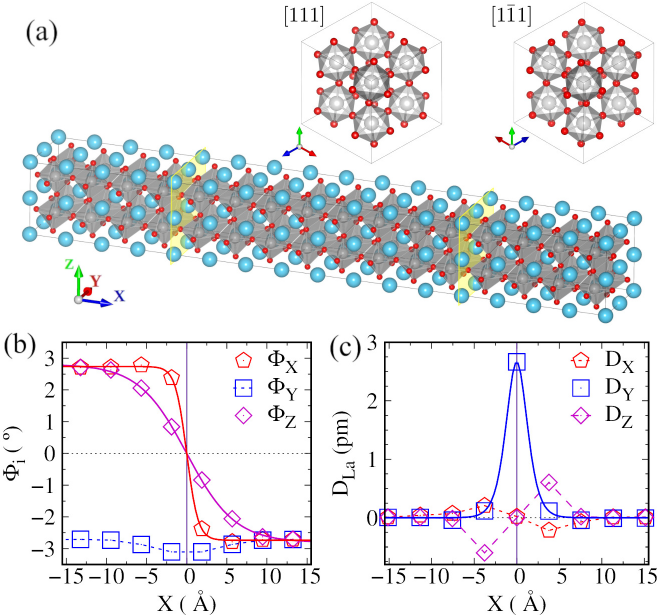}
		\caption{\label{figure_dft_SI} (a) Supercell  used for simulating two symmetric domains and DWs with periodic boundary conditions, comprising 64 5-atom pseudocubic cells for a total of 320 atoms. The DW planes are yellow colored, while above the two different domain states we display a top view of the tilting pattern along the corresponding rotation axes $[111]$ and $[1\bar{1}1]$. (b) Evolution of the tilting order parameter across a DW separating $[1\bar{1}1]$ and $[111]$ domains, locally realizing a $b^+a^-c^+$ tilting pattern. (c) Corresponding offcentering of La atoms, occurring mostly at the DW along the $[010]$ direction. Solid lines in (b) and (c) show the primary and secondary order parameters, described by $\Phi_0\tanh(X/W)$ and $D_0\sech(X/W)^{-2}$, respectively.}
\end{figure}

First-principles calculations have been performed within Density Functional Theory using the projector augmented wave (PAW) method \cite{PAW94,PAW_vasp99} as implemented in VASP \cite{kresse96a,kresse96b}. We adopted the generalized gradient approximation of Perdew, Burke and Ernzerhof (PBE) for the exchange-correlation functional \cite{Perdew96}. We considered a fully optimized structure of bulk rhombohedral LAO ($R\bar{3}c$ space group) with pseudocubic lattice parameters $a_{PC}=3.78$\AA~ and $\alpha_{PC}=90.19^\circ$, corresponding to lattice vectors $a_H=5.35$\AA~ and $c_H=13.05$\AA~ in hexagonal setting, consistently with experimental data of Ref. \cite{Hayward_2005}. We then constructed a supercell comprising 16 pseudocubic 5-atom unit cells along X direction and 2 unit cells along Y, Z directions, displayed in Figure \ref{figure_dft_SI}: such large supercells are required both to account for the tilting pattern of oxygen octahedra of LAO perovskite structure and because two DWs are needed to implement periodic boundary conditions. The rhombohedral phase of LAO is denoted as $a^-a^-a^-$ in Glazer notation \cite{Glazer1975}, where repeated letter $a$ indicates that the magnitude of O$_6$ rotation about each PC axis is the same and the negative superscript indicates an antiphase modulation of the rotations along that axis. Correspondingly, four possible domain states can be realized in the rhombohedral phase of LAO. They can be distinguished by the direction of the rotation axis around which the AlO$_6$ octahedra tilt, which is parallel to one of the four space diagonal of the (pseudo)cubic unit cell, i.e., $[111]$, $[1\bar{1}1]$, $[\bar{1}11]$ and $[11\bar{1}]$ in units of cubic axes describing the high-temperature, high-symmetry cubic LAO. Within each supercell, we accommodated two distinct domains obtained through a twin rotation operation, as shown in Fig.\ \ref{figure_dft_SI}(a), and then we optimized atomic positions until forces were smaller than 0.01 eV/\AA. We used a plane-wave cutoff of 500 eV and a 1$\times$4$\times$4 Monkhorst-Pack mesh for Brillouin-zone integration. We considered two possible DW realizations as representative cases, namely the DW between a $[111]$ domain and a $[\bar{1}11]$ (Fig.\ \ref{figure_dft} of the main text) or a $[\bar{1}1\bar{1}]$ (Fig.\ \ref{figure_dft_SI}) domain. The order parameter of the tilted structure can be constructed from the static rotational momenta of AlO$_6$ octahedra as \cite{litvin2011,iniguez2013,barone2014}:
\begin{align}
\bm \Phi(\bm R) = (-1)^{i+j+k}\,\sum_{l=1,6}\hat{\bm r}_l\times\hat{\bm r}^\prime_l
\end{align}
where $i,j,k$ are integer numbers and the lattice vector $\bm R=i\bm a_X+j\bm a_Y+k\bm a_Z$ describes the B-site position in the supercell. Unit vector $\hat{\bm r}_l$ ($\hat{\bm r}^\prime_l$) represents oxygen positions measured from each AlO$_6$ octahedron center of mass before (after) the static rotation about an axis passing through it. We verified that the O$_6$ center of mass coincides with the B-site sublattice within numerical accuracy.
The direction of the axial vector $\bm \Phi(\bm R)$ provides the axis about which the oxygen octahedron rotates and its magnitude yields
the rotation angle. The tilting pattern $a^-a^-a^-$ in the $[111]$ domain state is then described by $\bm \Phi=\Phi_0(\bm a_X+\bm a_Y+\bm a_Z)$, with a rotation angle of $\sqrt{3}\Phi_0$ around the $[111]$ axis. Similarly, the local offcentering of each La cation is evaluated from the center of mass of the LaO$_{12}$ dodecahedral oxygen cage as
\begin{align}
\textbf D_\text{La}(\bm R) =\frac{1}{12}\sum_{l=1,12}\,(\bm r^\text{La}-\bm r^\text{O}_l)
\end{align}
where $\bm r^\text{La}$ denotes the position of La atom within the unit cell identified by the vector lattice $\bm R$ and $\bm r^\text{O}_l$ represents the position of the first oxygen-coordinated dodecahedral shell of La. Both the tilting (primary) order parameter and A-site (secondary) offcentering are then layer-averaged in the $YZ$ plane to deliver their profile across the DWs \cite{barone2014}. The local tilting pattern at the DW between $[111]$ and $[\bar{1}11]$ domain states can be denoted as $c^+a^-a^-$, leading to a polar displacement of La cations along the [011] PC direction \cite{iniguez2013}, as shown in Figure \ref{figure_dft}(a),(b). Instead the DW between $[111]$ and $[\bar{1}1\bar{1}]$ domain states is described by $b^+a^-c^+$, leading to a La offcentering along the [010] direction \cite{iniguez2013}, as indeed found in our DFT simulations and shown in Figure \ref{figure_dft_SI}(b,c).

 \begin{figure*}[t]
		\includegraphics[width=0.75\textwidth]{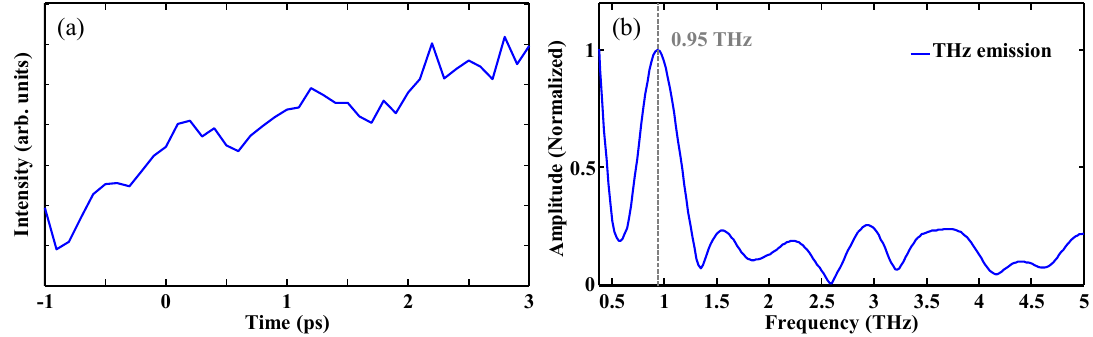}
		\caption{\label{Fig5} Measured THz radiation from the 50 $\mu$m thick LAO crystal excited by a 1030 nm pump pulse with 20 $\mu$J pulse energy and detected by the EOS method. (a) The measured time-domain THz waveform, and (b) the corresponding THz spectrum.}
\end{figure*}

The vibrational IR spectrum has been calculated within Density Functional Perturbation Theory (DFPT) \cite{baroni_2001} as implemented in Quantum ESPRESSO \cite{QE1,QE2}. We adopted PBE exchange-correlation functional and used PAW pseudopotentials taken from the Quantum ESPRESSO pseudopotential data base.  Given the computational load of simulating the IR vibrational spectra of the 320-atom supercell, we performed lattice-dynamics calculations for the rhombohedral bulk phase of LAO. For DFPT calculations we adopted a stringent kinetic energy cutoff of 80 Ry and a charge-density cutoff of 580 Ry, using a 15$\times$15$\times$15 Monkhorst-Pack mesh for Brillouin-zone integration. We reoptimized the internal degrees of freedom to a precision of 0.0025~eV/\AA, and consider two artificially distorted structures where the La and Al sublattices are displaced along the $[011]$ and $[010]$ PC directions by 2.5 pm. After the oxygen positions have been further refined, we calculated the IR vibrational intensity of a phonon mode $\mu$ with frequency $\omega_\mu$ as:
\begin{align}
I_\mu =\sum_\alpha \left\vert \sum_{s,\beta}\frac{1}{\sqrt{M_s}} Z^{s}_{\alpha\beta}e^{s}_{\mu,\beta} \right\vert^2
\end{align}
where $M_s$ and $\bm Z^s$ are, respectively, the mass and the  Born effective charge tensor of atom $s$, while $\bm e^s_\mu$ is the phonon polarization vector. The IR spectra shown in Fig.\ \ref{figure_dft}(d) of the main text have been broadened using a Lorentzian function for each calculated phonon mode, centered at the corresponding $\omega_\mu$ and  with phenomenological damping rate $\gamma_\mu/2\pi=0.1$ THz.
\section{THz emission}

To further support our interpretation, we measured the THz emission from the LAO crystal, which exhibits a very weak but measurable narrowband THz spectrum with a peak at $\sim$ 1~THz (see Fig.~\ref{Fig5}). Here, the LAO crystal is excited by a 1030 nm pump pulse with 20 µJ pulse energy, and the emitted THz field is detected by the EOS method. This narrowband THz emission can be explained by the Raman-like excitation of the $E_g$ phonon mode, that then emits a THz field at the phonon frequency. This observation provides experimental evidence for the IR-activity of the $E_g$ mode, thereby supporting our interpretation of the observed cross-peak C and C$^\prime$ as consequence of anharmonic couplings of the $E_g$ phonon with the Raman-active $A_{1g}$ mode. 
%
\end{document}